# High performance photonic microwave filters based on a 50GHz optical soliton crystal Kerr micro-comb†

Xingyuan Xu,[1] Mengxi Tan,[1] Jiayang Wu,[1] *Member, IEEE,* Thach G. Nguyen,[2] Sai T. Chu,[3] Brent E. Little,[4] Roberto Morandotti,[5] *Senior Member, IEEE,* Arnan Mitchell,[2] *Member, IEEE,* and David J. Moss,[1] *Fellow, IEEE*

[1] X. Xu, M. Tan, J. Wu, and D. J. Moss are with Centre for Micro-Photonics, Swinburne University of Technology, Hawthorn, VIC 3122, Australia. (Corresponding e-mail: dmoss@swin.edu.au).

[2]T. G. Thach and A. Mitchell are with the School of Engineering, RMIT University, Melbourne, VIC 3001, Australia.

[3]S. T. Chu is with Department of Physics and Material Science, City University of Hong Kong, Tat Chee Avenue, Hong Kong, China.

[4]B. E. Little is with State Key Laboratory of Transient Optics and Photonics, Xi'an Institute of Optics and Precision Mechanics, Chinese Academy of Science, Xi'an, China.

[5]R. Morandotti is with INSR-Énergie, Matériaux et Télécommunications, 1650 Boulevard Lionel-Boulet, Varennes, Québec, J3X 1S2, Canada, with ITMO University, St. Petersburg, Russia, and also with Institute of Fundamental and Frontier Sciences, University of Electronic Science and Technology of China, Chengdu 610054, China.

***Abstract*—We demonstrate a photonic radio frequency (RF) transversal filter based on an integrated optical micro-comb source featuring a record low free spectral range of 49 GHz yielding 80 micro-comb lines across the C-band. This record-high number of taps, or wavelengths for the transversal filter results in significantly increased performance including a $Q_{RF}$ factor more than four times higher than previous results. Further, by employing both positive and negative taps, an improved out-of-band rejection of up to 48.9 dB is demonstrated using Gaussian apodization, together with a tunable centre frequency covering the RF spectra range, with a widely tunable 3-dB bandwidth and versatile dynamically adjustable filter shapes. Our experimental results match well with theory, showing that our transversal filter is a competitive solution to implement advanced adaptive RF filters with broad operational bandwidths, high frequency selectivity, high reconfigurability, and potentially reduced cost and footprint. This approach is promising for applications in modern radar and communications systems.**

## I. INTRODUCTION

Radio frequency (RF) filters are key components in modern radar and communications systems [1]. Driven by the demand for improved performance, advanced RF filters that have broad operational bandwidths, high reconfigurability, high frequency selectivity, and low cost are highly desired [2-6]. Whereas electrical RF filters suffer from the intrinsic electronic bandwidth bottleneck, photonic RF filters can provide much higher RF bandwidths, highly reconfigurable filter shapes, as well as rapid tunability and strong immunity to electromagnetic interference.

There are a number of approaches to realizing photonic RF filters, including ones that map the optical filter response onto the RF domain [7-12], highlighted by on-chip (waveguide based) stimulated Brillouin scattering [10-12]. This approach has achieved extremely high performance in terms of RF resolution — as high as 32 MHz — and a stopband rejection >55 dB. Another key approach focuses on reconfigurable transfer functions for adaptive signal processing based on transversal filters [13-18]. These operate by generating weighted and progressively delayed replicas of the RF signal in the optical domain and then combining them upon photo-detection. Transverse filters are capable of achieving arbitrary RF transfer functions by simply changing the tap weights, and so are attractive for the implementation of advanced adaptive and dynamic RF filters. Typically, discrete laser arrays [19-21] or Bragg grating arrays [22, 23] have been employed to supply the needed taps. However, while offering many advantages, these approaches result in significantly increased complexity as well as reduced performance due to the limited number of available taps. Alternative approaches, including those based on optical frequency comb sources achieved by electro-optic (EO) modulation, can help mitigate this problem, but they require cascaded high frequency EO [24-26] or Fabry-Perot EO [27] modulators that in turn require high-frequency RF sources. By far the most attractive approaches are based on integrated photonics.

All-optical signal processing based on silicon [28, 29] has been extremely successful for all-optical logic [30], ultra-high speed demultiplexing [31, 32], optical performance monitoring [33, 34], regeneration [35, 36], and others [37-43]. Since CMOS (complementary metal oxide semiconductor) compatible platforms are centrosymmetric, nonlinear devices have been based on third order nonlinearities - the Kerr nonlinearity ($n_2$) [28, 29] and third harmonic generation [38, 44-48]. However, while silicon has an extremely high nonlinearity $\gamma = \omega n_2 / c A_{eff}$, it also has high two-photon absorption (TPA, β) and a poor nonlinear figure of merit of 0.3 (FOM = $n_2$ / (β λ)) in the telecom band. While TPA can be advantageous [49-51], it is generally a limitation and this inspired interest in other platforms including chalcogenide glasses [52-61]. In 2008 new nonlinear CMOS platforms [62-74] enabled the first integrated micro-combs [63, 64] following the discovery of Kerr combs in 2007 [75]. Many breakthroughs have since been reported including mode-locked lasers [76-79], quantum physics [80-86], optical frequency synthesis [87], ultrahigh bandwidth communications [88], and others [89-95]. The success of these platforms and others such as a-Si [96] arises from their low linear loss, high nonlinearity, and very low TPA.

Kerr micro-comb sources [63-65, 74-79, 86-95, 97, 98], particularly those based on CMOS-compatible platforms [63-65], offer distinct advantages over traditional multi-wavelength sources for transversal RF filters, such as the potential to provide a much higher number of wavelengths as well as greatly reduced footprint and complexity. RF transversal filters based on Kerr micro-combs [99-105] have achieved a high degree of versatility and dynamic reconfigurability. However, to date the relatively large comb spacing arising from the large free spectral range (FSR) of ~1.6 nm (~200 GHz) has restricted the number of taps to less than 21 within the 30nm wide bandwidth of the C-band. This is an important consideration since this approach requires optical amplifiers and optical spectral shapers which are more readily available for wavelengths in the C-band. The limitation in the number of taps has limited the performance of micro-comb based transversal RF filters in terms of frequency selectivity, tuning resolution, dynamic versatility (filter shapes), as well as the performance of RF photonic signal processors [100-105].

In this paper, we demonstrate a micro-comb-based photonic RF transversal filter with a record high number of taps, featuring 80 wavelengths over the C-band. This is the highest number so far reported for micro-comb-based RF transversal filters, enabled by a 49GHz-free-spectral-range integrated Kerr micro-comb source. This resulted in a $Q_{RF}$ factor for the RF bandpass filter of four times higher than previous results [100]. Further, by programming and shaping the Kerr optical micro-comb, we achieve RF filters with a high out-of-band rejection of up to 48.9 dB using Gaussian apodization, as well as a significantly improved tunable centre frequency covering the RF spectra range (from $0.05\times FSR_{RF}$ to $0.40\times FSR_{RF}$). We also demonstrate an adaptive photonic RF filter with highly reconfigurable 3-dB bandwidths (from 0.5 to 4.6 GHz) and arbitrary filter shapes. Our experimental results agree well with theory, verifying the feasibility of our approach towards the realization of high performance advanced adaptive RF transversal filters with potentially reduced cost, footprint, and complexity than other solutions.

## II. Theory

The transfer function of a photonic transversal filter can be described as

$$H(\omega)=\sum_{n=0}^{N-1}h(n)e^{-j\omega nT} \tag{1}$$

where $\omega$ is the angular frequency of the input RF signal, $N$ is the number of taps, $h(n)$ is the discrete impulse response representing the tap coefficient of the $n_{th}$ tap, and $T$ is the time delay between adjacent taps. The free spectral range of the transversal filter $FSR_{RF}$ is given by $1/T$. By properly setting the tap coefficients ($h(n)$, $n$ = 0, 1, …, $N$-1) for different spectral transfer functions, a reconfigurable transversal filter with arbitrary filter shapes can be achieved [1].

Figure 1 shows a schematic diagram of the 80-tap photonic RF transversal filter based on an integrated Kerr micro-comb source. The Kerr optical frequency combs were generated in an integrated microring resonator (MRR) pumped by a continuous-wave (CW) laser, amplified by an erbium-doped fibre amplifier, with the polarization adjusted via a polarization controller to optimize the power coupled to the MRR. When the pump wavelength was swept across one of the MRR's resonances and the pump power was high enough to provide sufficient parametric gain, optical parametric oscillation occurred, ultimately generating Kerr optical combs with a spacing equal to the free spectral range of the MRR. The generated Kerr micro-comb served as a multi-wavelength source where the power of each comb line was manipulated by the Waveshapers to achieve the designed tap weights.

The shaped comb lines were then fed into an EO intensity modulator, yielding replicas of the input RF signal in the optical domain. The modulated signal produced by the intensity modulator went through a spool of dispersive fiber, generating a time delay $T$ between adjacent taps. Finally, the weighted and delayed taps were combined upon photo detection and converted back into RF signals at the output.

## III. Experimental results

The MRR used to generate the Kerr optical micro-comb (Fig. 2(a)) was fabricated on a high-index doped silica glass platform using CMOS-compatible fabrication processes [65]. First, high-index (n = ~1.7 at 1550 nm) doped silica glass films were deposited using plasma enhanced chemical vapour deposition, then patterned by deep ultraviolet photolithography and etched via reactive ion etching to form waveguides with exceptionally low surface roughness. Finally, silica (n = ~1.44 at 1550 nm) was deposited as an upper cladding. The advantages of our platform for optical micro-comb generation include ultra-low linear loss (~0.06 $dB\cdot cm^{-1}$), a moderate nonlinear parameter (~233 $W^{-1}\cdot km^{-1}$), and in particular a negligible nonlinear loss up to extremely high intensities (~25 $GW\cdot cm^{-2}$). Due to the ultra-low loss of our platform, the MRR features narrow resonance linewidths [105], corresponding to a quality factor of ~1.5 million (Fig. 2(b)). After packaging the device with fiber pigtails, the through-port insertion loss was ~1 dB, assisted by on-chip mode converters. The radius of the MRR was ~592 μm, corresponding to an optical free spectral range of ~0.4 nm or ~ 49 GHz (Fig. 2(c)). The record small optical free spectral range of the MRR enabled 80 wavelengths, or taps for the transversal filter, in the C band—much larger than that used in previous reports [100,101].

References done to here

To generate Kerr micro-combs, we adjusted the polarization and wavelength of the pump light to one of the TE resonances of the MRR at ~1553.2 nm [102]. The pump power was set at ~30.5 dBm. When the detuning between the pump wavelength and the cold resonance became small enough, such that the intra-cavity power reached a threshold value, modulation instability (MI) driven oscillation was initiated [74]. Primary combs were thus generated with the spacing determined by the MI gain peak — mainly a function of the intra-cavity power and dispersion. As the detuning was changed further, distinctive 'fingerprint' optical spectra were observed (Fig. 3(a)). These spectra are similar to what has been reported from spectral interference between tightly packed solitons in the cavity — so called "soliton crystals" [97, 98]. An abrupt step in the measured intracavity power (Fig. 3(b)) was observed at the point where these spectra appeared, as well as a dramatic reduction in the RF intensity noise (Fig. 3(c)). Together we take these observations as indicative of possible soliton crystal formation, although to conclusively demonstrate this one would need to perform time resolved pulse autocorrelation measurements, which we did not do. In any case, the key issue for our experiments, however, was not the specific nature of the micro-comb state of oscillation so much as the low RF noise and high coherence, which were relatively straightforward to achieve through adiabatic pump wavelength sweeping. We found that it was not necessary to specifically achieve any specific state, including either soliton crystals or single soliton states in order to obtain high performance — only that the chaotic regime [74] should be avoided. This is important since there are a much wider range of coherent low RF noise states that are more readily accessible than any specific soliton related state. For convenience, in this paper we refer to the micro-comb states that we generate as being soliton crystal states, bearing in mind the above caveats.

The soliton crystal comb was then spectrally shaped via a two-stage optical spectral shaper (Finisar, Waveshaper 4000S) in order to enable a larger dynamic range of loss control and higher shaping accuracy than a single stage [99, 100]. The micro-comb was first pre-shaped to reduce the power difference between the comb lines to less than 15 dB, and then amplified and accurately shaped by a subsequent Waveshaper according to the designed tap weights. For each Waveshaper, a feedback control path was adopted to increase the accuracy of the comb shaping, where the power in the comb lines was detected by an optical spectrum analyzer and compared with the ideal tap weights in order to generate error signals for calibration.

The shaped comb lines were divided into two parts by the Waveshaper according to the algebraic sign of the tap coefficients, and then fed into the 2×2 balanced Mach-Zehnder modulator (MZM) biased at quadrature. The comb lines corresponding to positive taps were modulated on the positive slope of the MZM transfer function, while the comb lines corresponding to negative taps were modulated on the negative slope, thus yielding replicas of the input RF signal with opposite phase and tap coefficients having opposite algebraic signs. Then the signal went through ~5-km of standard single mode fibre (SMF) to provide the progressive tap delays. The dispersion of the SMF was ~17.4 ps / (nm · km), corresponding to a time delay $T$ of ~34.8 ps between adjacent taps, yielding an operation bandwidth (i.e., the Nyquist frequency, half of $FSR_{RF}$) of ~14.36 GHz for the transversal filter. This operation bandwidth can be easily enlarged by decreasing the time delay (e.g., using a shorter spool of SMF), at the expense of a reduced tuning resolution. However, the maximum operational bandwidth of the transversal filter is limited by the comb spacing. Significant crosstalk between adjacent wavelength channels (or taps) occurs for RF operation beyond 24.5 GHz — half of the microcomb's spacing 49 GHz. This issue can be addressed by employing a microcomb source with a larger comb spacing, although at the expense of providing fewer comb lines/taps across the C-band. Finally, the weighted and delayed taps were combined and converted back into RF domain via a high-speed photodetector (Finisar, 40 GHz bandwidth).

To demonstrate the enhancement in the frequency selectivity, or resolution of the transversal filters, brought about by the large number of taps and represented by the $Q_{RF}$ factor, we shaped the micro-comb to implement a low-pass sinc filter featuring equal tap weights (i.e., $h_{sinc}(n)=1$) for different numbers of taps. We then measured the 3-dB bandwidth ($BW_{sinc}$) to calculate the corresponding $Q_{RF}$ factor ($Q_{RF} = BW_{sinc} / FSR_{RF}$). In the experiments we selected different numbers of taps ranging from 2 to 80. The corresponding optical spectra are shown in Fig. 4. The RF transmission spectra of the sinc filter (Fig. 5(a)) were measured by a vector network analyser (VNA, Anritsu 37369A), and showed good agreement with theory (Fig. 5(b)). The measured $BW_{sinc}$ decreased from 3.962 to 0.236 GHz when the tap number was increased from 2 to 80, indicating a greatly enhanced $Q_{RF}$ of up to 73.7 with 80 taps — four times larger than previous demonstrations [100]. The theoretical $BW_{sinc}$ and $Q_{RF}$ as a function of the tap number are shown in Fig. 5(c). As can be seen, $Q_{RF}$ increases linearly with the tap number, further confirming the significant improvement of the frequency selectivity or resolution (reflected by $Q_{RF}$) brought about by the large tap number used here.

To improve the performance of the transversal filter in terms of out-of-band rejection, a Gaussian apodization was applied to the sinc filter [17], which was achieved by modifying the original discrete impulse response (i.e., tap weights) to the product of itself and a Gaussian function, given by

$$h_{Gau}(n) = h_{\sin c}(n) \cdot e^{-\frac{(n-40.5)^2}{2\sigma^2}} \tag{2}$$

where $\sigma$ is the root mean square width of the Gaussian function. Figure 6 shows the shaped comb spectra and Fig. 7 shows the corresponding RF transmission spectra. As a function of decreasing $\sigma$, the main-to-secondary sidelobe ratio (MSSR) of the sinc filter increased from 26.4 to 48.9 dB, although at the expense of a deteriorated $Q_{RF}$ factor. While illustrating the capability of our transversal filter to achieve high out-of-band rejection, it also indicates that a large number of taps would be needed if both a very high $Q_{RF}$ factor and MSSR are simultaneously required. With the ability to offer a large number of taps — potentially over 200 in the C+L bands, micro-combs serve as a highly attractive approach to meet these demands. We note that the experimental results for $\sigma = 10$ were limited by our measurement noise floor of the VNA, and that up to 74.5 dB can in principle be achieved.

In order to demonstrate the tunability of the 80-tap transversal filter's centre frequency, the tap coefficients of the Gaussian-apodized sinc filter (we set σ =16 to obtain an acceptable MSSR) were multiplied by a sine function to shift the RF transmission spectrum [18]. The corresponding discrete impulse response is given by

$$h_{TCF}(n) = h_{Gau}(n) \cdot \cos\frac{f_{centre} \cdot \pi n}{FSR_{RF}} \tag{3}$$

where $f_{centre}$ is the tunable centre frequency. Figure 8 shows the shaped comb spectra of the centre-frequency-tunable transversal filter. The corresponding RF transmission spectra in Fig. 9 shows a tunable centre frequency ranging from $0.05\times FSR_{RF}$ =1.4 GHz to $0.40\times FSR_{RF}$ =11.5 GHz with a relatively high MSSR of >25 dB. The third-order dispersion (TOD) of the SMF (~0.083 ps/nm$^2$/km) was taken into consideration during the calculation of the RF transmission spectra.

Next, we demonstrated the high reconfigurability of the 80-tap transversal filter in terms of bandwidth and filter shapes using a straight-forward approach. First, $M$ sets of tap coefficients corresponding to bandpass filters with progressive centre frequencies were calculated (with the method described in Eq. (3)), given as $h_{TCF}(k,n)$, $k$=1, 2, 3, …, $M$. Secondly, the calculated tap coefficients were weighted according to the designed filter shapes, with the weights given by $w(k)$. Finally, the weighted tap coefficients were added up together and the constructed discrete

impulse response was

$$h_{total}(n) = \sum_{k=1}^{M} h_{TCF}(k,n) \cdot w(k) \tag{4}$$

To verify the ability of our approach to generate arbitrary filter shapes, we set $w(k)=k$ and $w(k)=M-k$ (both with $M=10$), respectively, to achieve positive and negative slopes for the RF transmission, as shown in Fig. 10 (b, d). In addition, we varied $M$ from 1 to 9 (with $w(k)=1$) to achieve a rectangular filter with tunable bandwidths ranging from 0.5 to 4.6 GHz (Fig. 10(f)). The passband flatness for the $M=9$ case was < 2.9 dB, with out-of-band rejection ratio > 15 dB and edge steepness (roll-off rate) >15 dB/GHz. Our record performance filters primarily resulted from the record number of taps (80) that we employed, enabled by the fine 49GHz spacing of our micro-comb.

Finally, considerable effort [1, 18] has been made on transversal filters that make use solely of positive taps. This approach has been used since it offers significantly reduced complexity. Therefore, we also performed experiments (and theory) based on this approach for comparison purposes. By including experiments based on all-positive tap coefficient transversal filters here, we demonstrate the significant advantages of being able to use both positive and negative taps on the filter performance.

For the positive tap experiments, the bandwidth of transversal filter was tuned by shifting the transmission spectrum of a low-pass filter with varying bandwidth (designed by Remez algorithm [107]) by half of $FSR_{RF}$, and the corresponding discrete impulse response is given by

$$h_{TBW}(n) = h_{LPF}(n) \cdot \cos \pi n - h_{\min} \tag{5}$$

where $h_{LPF}(n)$ is the discrete impulse response of the low-pass filter, $h_{min}$ is the minimum tap weights among $h_{LPF}(n)\cdot\cos\pi n$, $n=1, 2, 3, \ldots, 80$ (e.g., $h_{min} = -0.2434$ for the 7.18GHz-bandwidth filter, Fig. 12, purple line). Figure 11 shows the shaped comb spectra of the all-positive taps tunable transversal bandpass filter which, in light of the feedback shaping control path, matched closely with the designed tap weights. The measured RF transmission spectra (Fig. 12, upper plot) shows a tunable bandwidth from 0.89 to 7.18 GHz and matches closely with theory (Fig. 12, lower plot) when taking the third-order dispersion (TOD) of the SMF (~0.083 ps/nm$^2$/km) into account. As can be seen, the all-positive-tap filter's transmission spectrum (Fig. 12) was the superposition of the target filter and a sinc filter, which introduced spurious passbands at the baseband [18]. Although these passbands at low frequencies can be suppressed by another highpass filter, there were still limitations imposed onto the overall MSSR of the transversal filter, which is a general feature present in all all-positive tap filters and is one of their main drawbacks [1]. This was a prime motivation for employing both positive and negative taps in this work. This work demonstrates that the soliton crystals used here have enormous potential for microwave and RF signal processing functions based on optical microcombs. [108-178]

## IV. Conclusion

We demonstrate record performance and versatility for microcomb-based photonic RF transversal filters by employing a 49GHz-FSR integrated optical micro-comb source capable of providing a record high number of wavelengths, or taps—80 over the C-band. We achieve $Q_{RF}$ factors that are four times larger than previously demonstrated, as well as a high out-of-band rejection of up to 48.9 dB using Gaussian apodization. We also achieve a tunable centre frequency covering the RF spectral range (from $0.05\times FSR_{RF}$ to $0.40\times FSR_{RF}$, or 1.4 to 11.5 GHz), as well as tunable 3dB bandwidths ranging from 0.5 to 4.6 GHz. Finally, we demonstrate highly reconfigurable filter shapes with positive and negative slopes across the passband as well as rectangular bandpass filter shapes. The experimental results agree well with theory, verifying that our transversal filter is a competitive approach towards achieving advanced adaptive RF transversal filters with broad operational bandwidths, high frequency selectivity, high reconfigurability, and potentially reduced cost and footprint, all of which are critical issues for modern radar and communications system.

## Acknowledgements

This work was supported by the Australian Research Council Discovery Projects Program (No. DP150104327). RM acknowledges support by the Natural Sciences and Engineering Research Council of Canada (NSERC) through the Strategic, Discovery and Acceleration Grants Schemes, by the MESI PSR-SIIRI Initiative in Quebec, and by the Canada Research Chair Program. He also acknowledges additional support by the Government of the Russian Federation through the ITMO Fellowship and Professorship Program (grant 074-U 01) and by the 1000 Talents Sichuan Program in China. Brent E. Little was supported by the Strategic Priority Research Program of the Chinese Academy of Sciences, Grant No. XDB24030000.

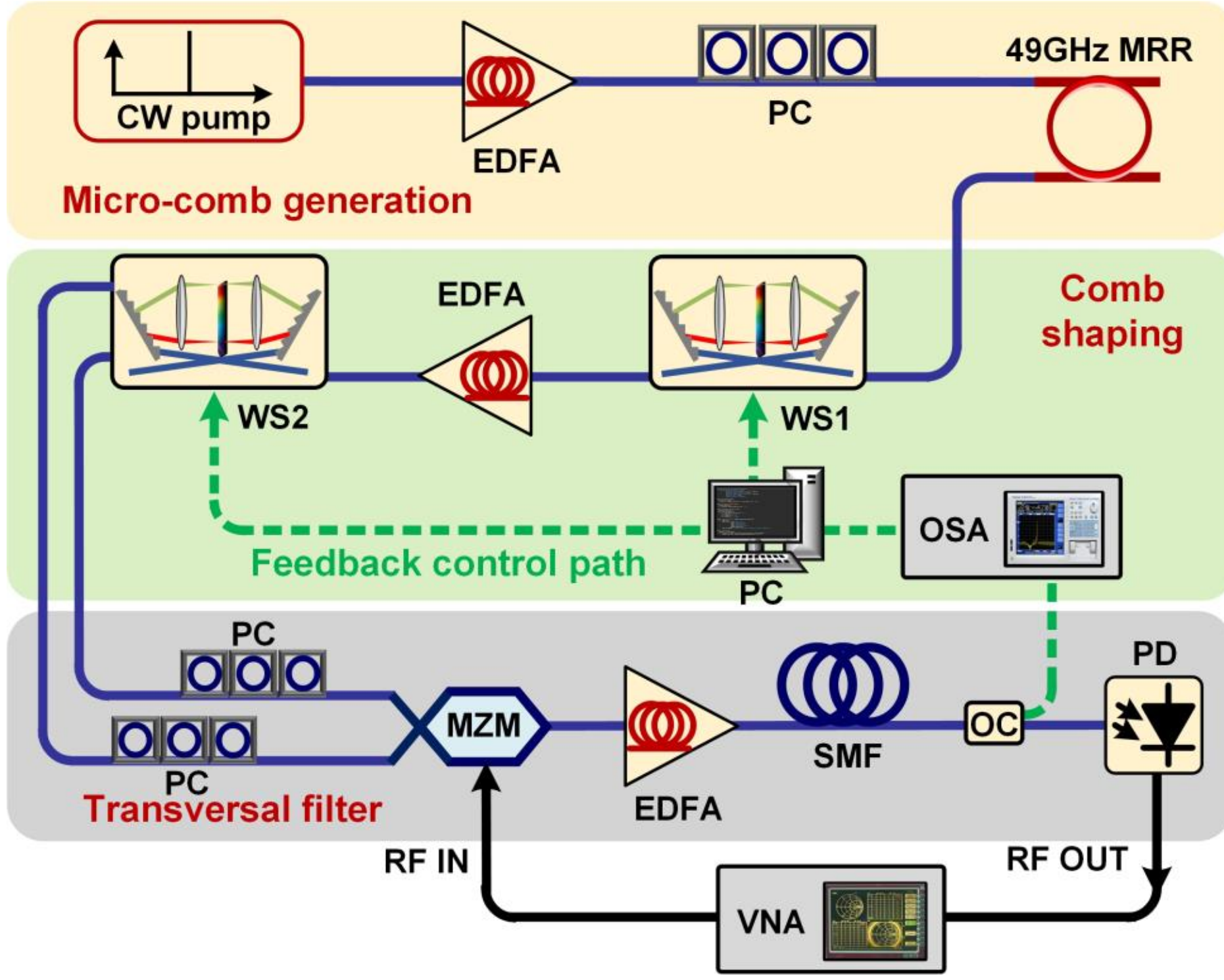

Fig. 1. Schematic diagram of the 80-tap photonic RF transversal filter based on an integrated 49GHz-spacing micro-comb source. EDFA: erbium-doped fiber amplifier. PC: polarization controller. MRR: micro-ring resonator. WS: Waveshaper. MZM: Mach-Zehnder modulator. SMF: single mode fiber. OC: optical coupler. OSA: optical spectrum analyzer. PC: computer. PD: photodetector. VNA: vector network analyzer.

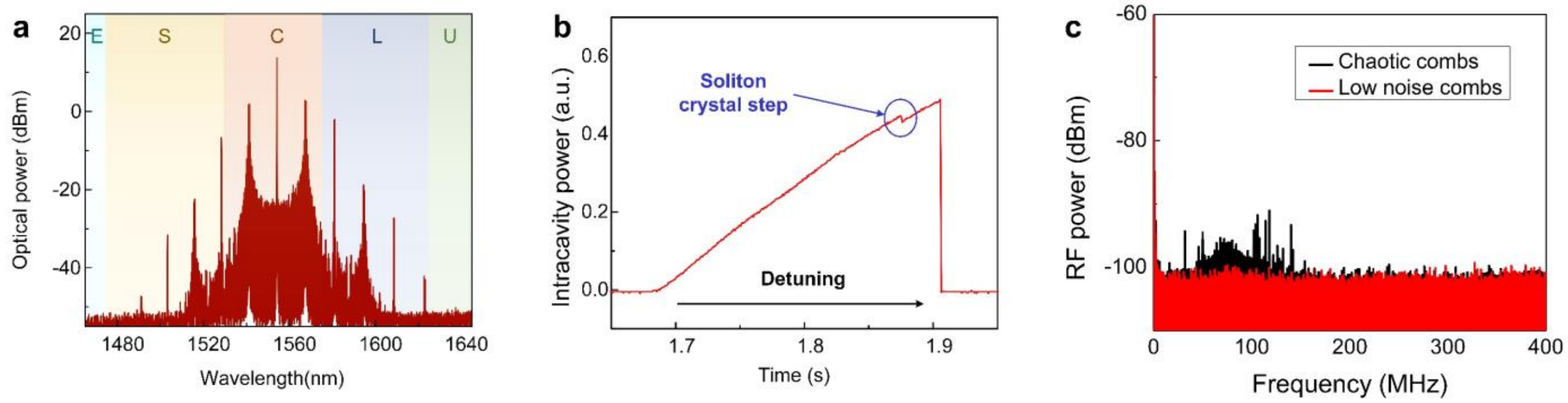

Fig. 3. (a) Optical spectrum of the generated micro-combs with 200 nm span. (b) Measured transmission of a single resonance showing the soliton crystal step, and (c) the measured RF spectra.

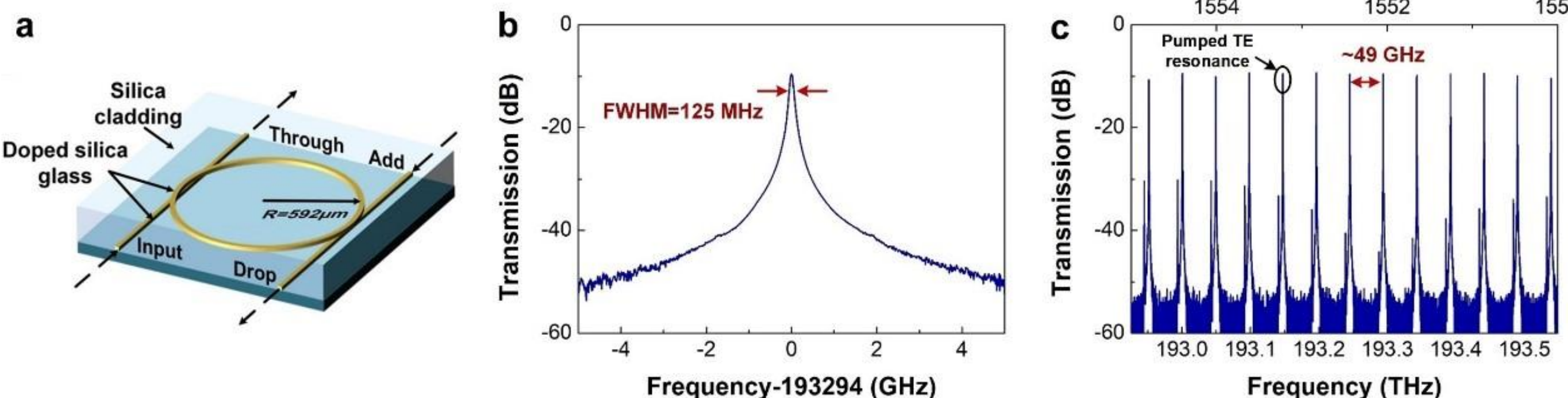

Fig. 2. (a) Schematic illustration of the integrated MRR for generating the Kerr micro-comb. (b) A resonance at 193.294 THz with full width at half maximum (FWHM) of 124.94 MHz, corresponding to a quality factor of $1.549\times10^{6}$. (c) Drop-port transmission spectrum of the integrated MRR with a span of 5 nm, showing an optical free spectral range of 49 GHz.

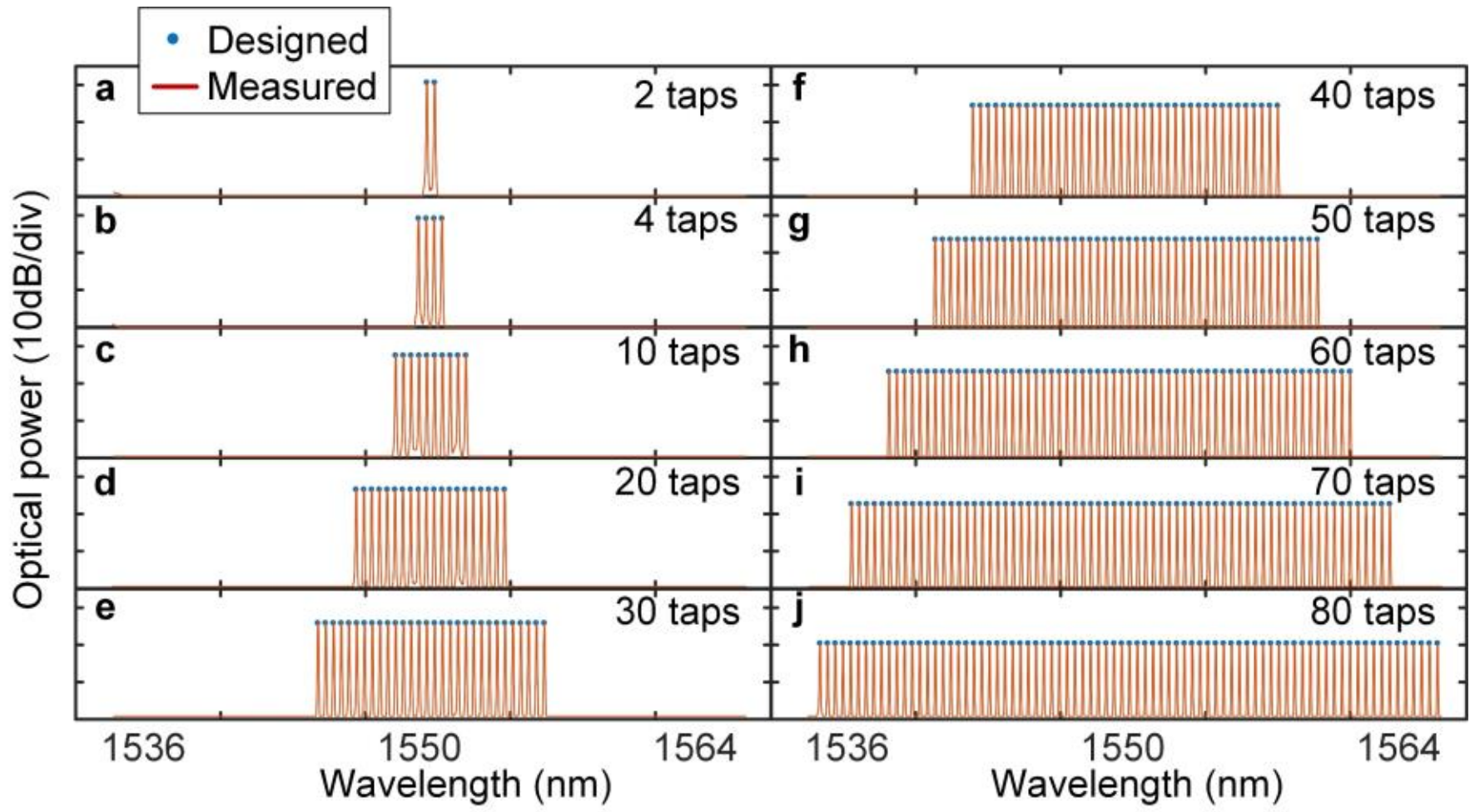

Fig. 4. Optical spectra of the shaped micro-comb corresponding to the tap weights of a sinc filter with different tap number.

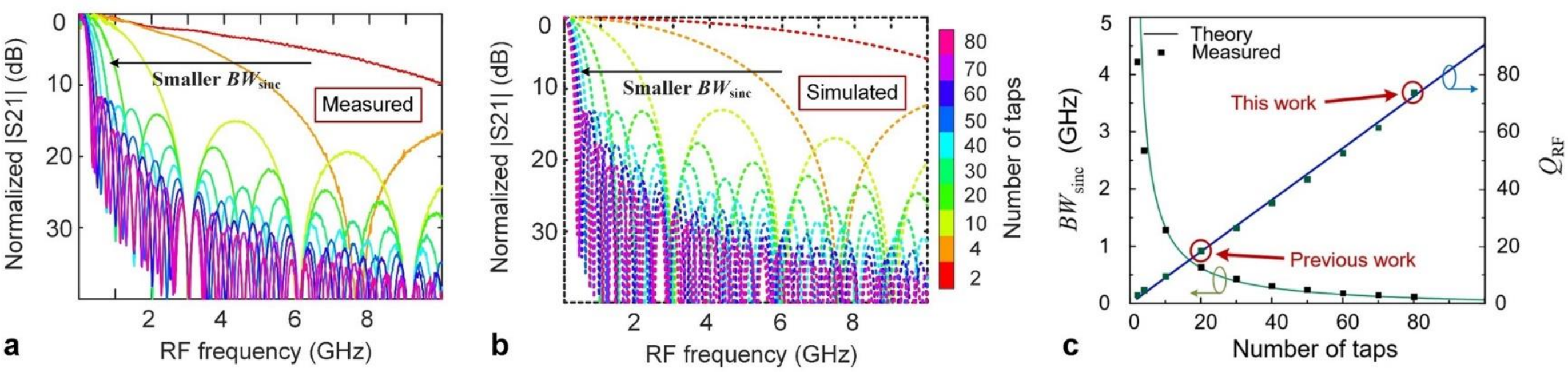

Fig. 5. (a) Measured and (b) simulated and RF transmission spectra of the sinc filter with different tap numbers, and (c) extracted corresponding 3-dB bandwidths ($BW_{sinc}$) and $Q_{RF}$ factors.

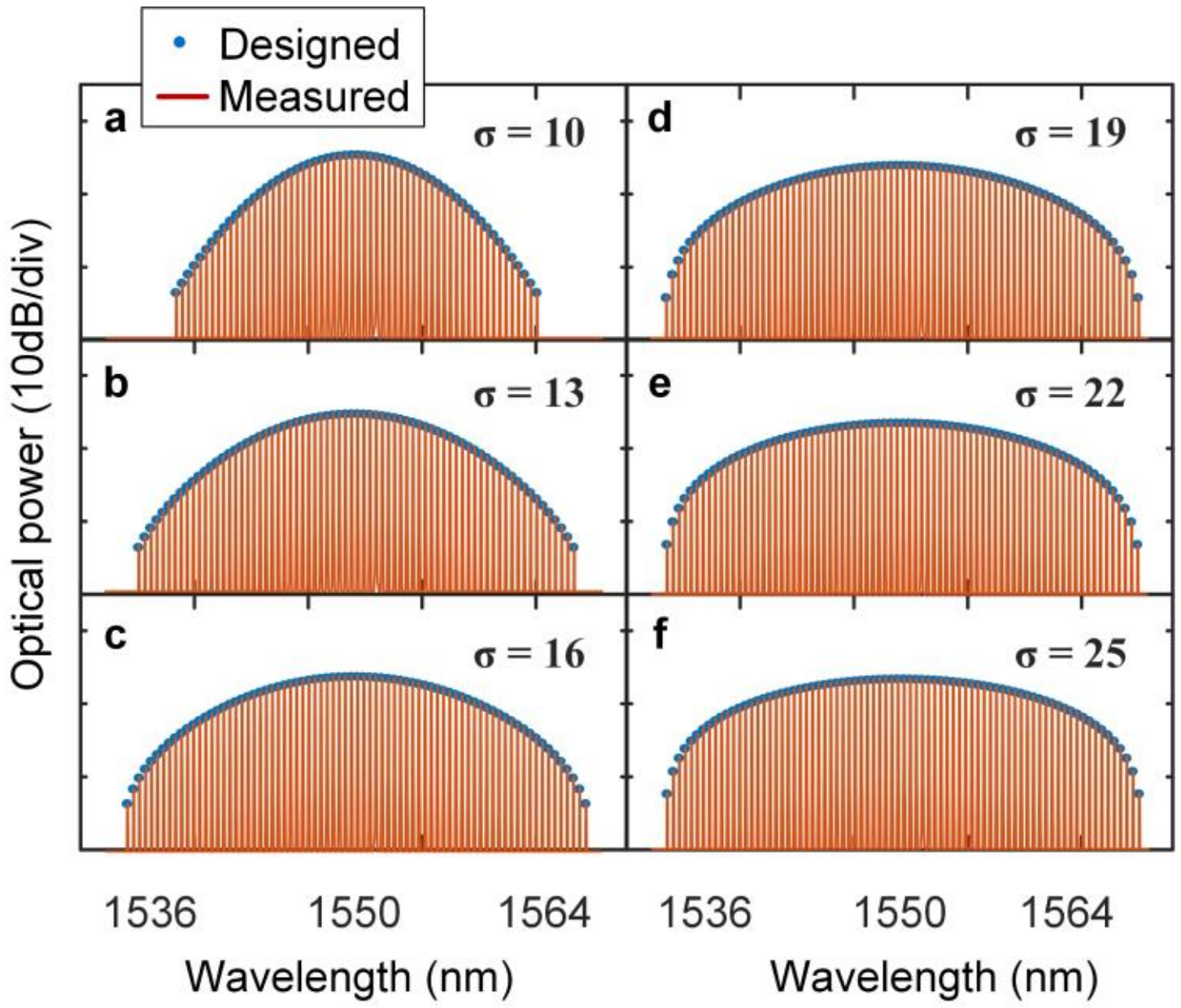

Fig. 6. Optical spectra of the shaped micro-comb corresponding to the Gaussian-apodized sinc filter. (a-f) corresponds to the root mean square width of the Gaussian function $\sigma$ set from 10 to 25, with a step of 3.

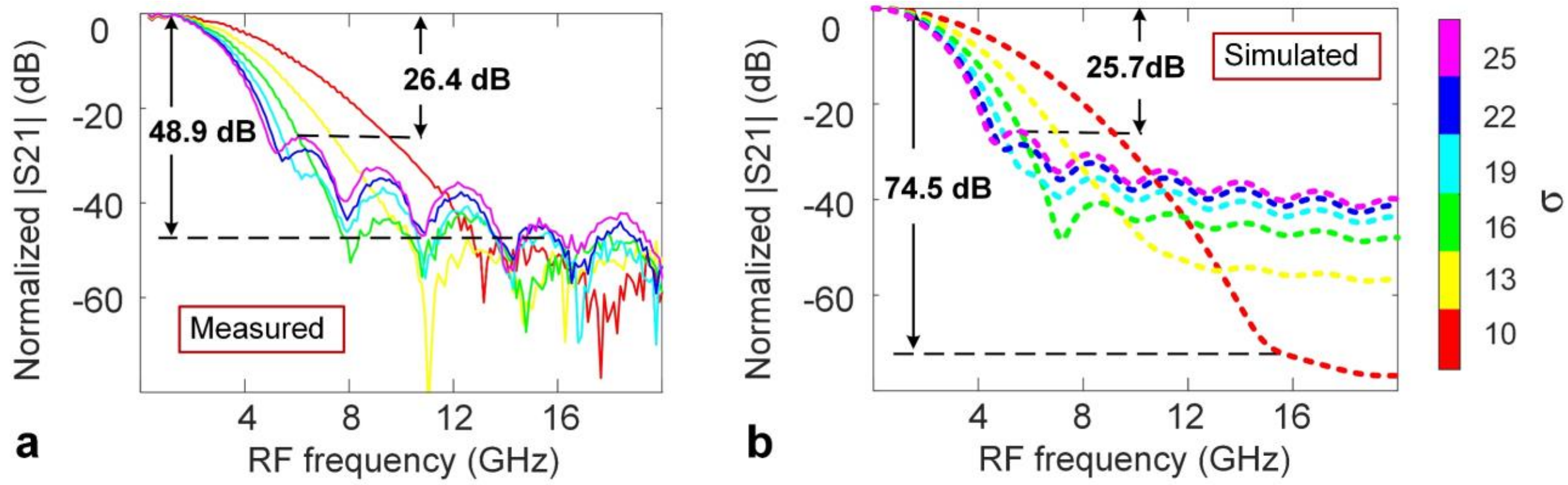

Fig. 7. (a) Measured and (b) simulated RF transmission spectra of the Gaussian-apodized sinc filter.

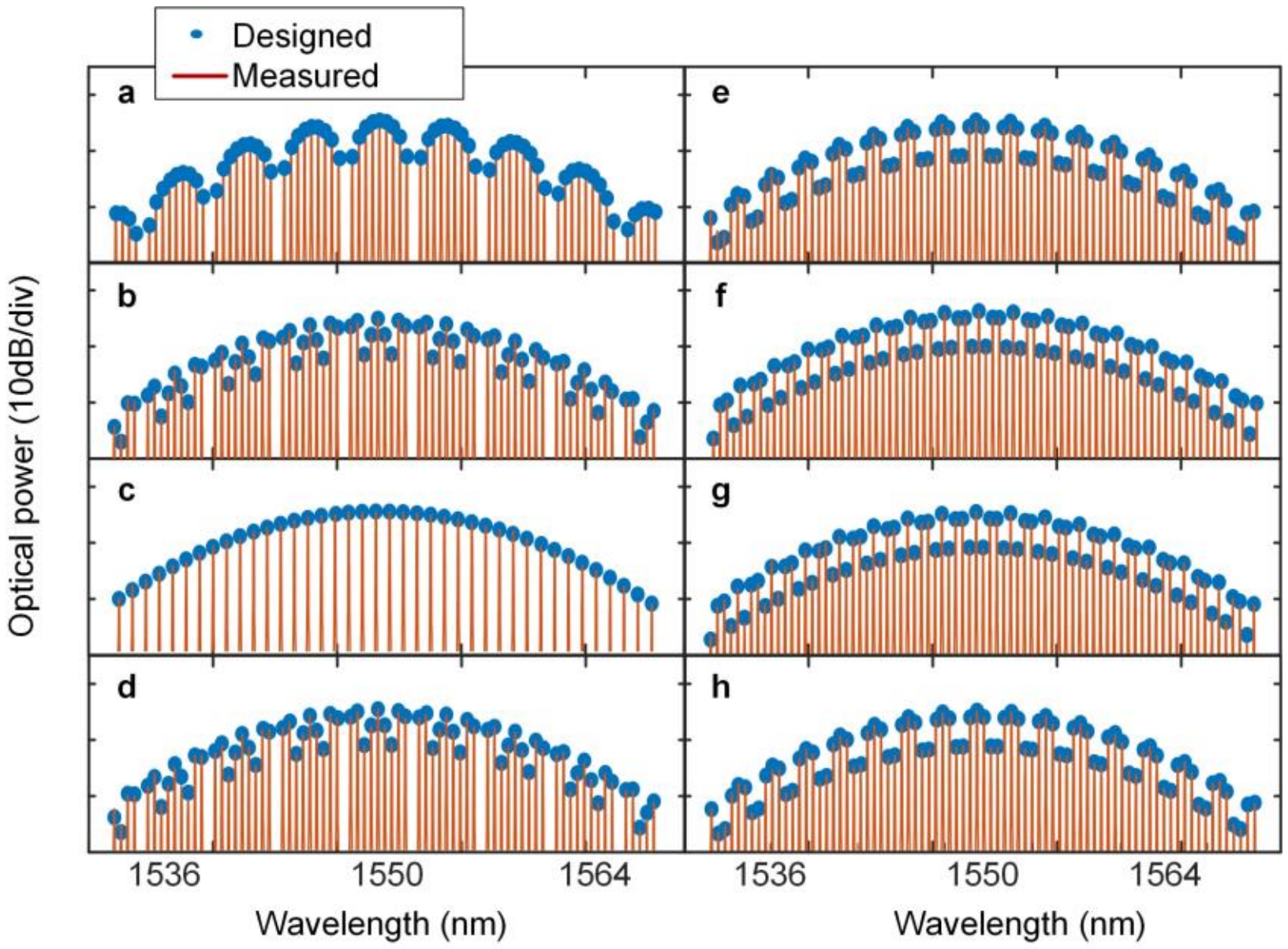

Fig. 8. Optical spectra of the shaped micro-comb corresponding to the centre-frequency-tunable Gaussian apodized sinc filter. (a-h) corresponds to a designed centre frequency ranging from $0.05\times FSR_{RF}$ to $0.40\times FSR_{RF}$, with a step of $0.05\times FSR_{RF}$.

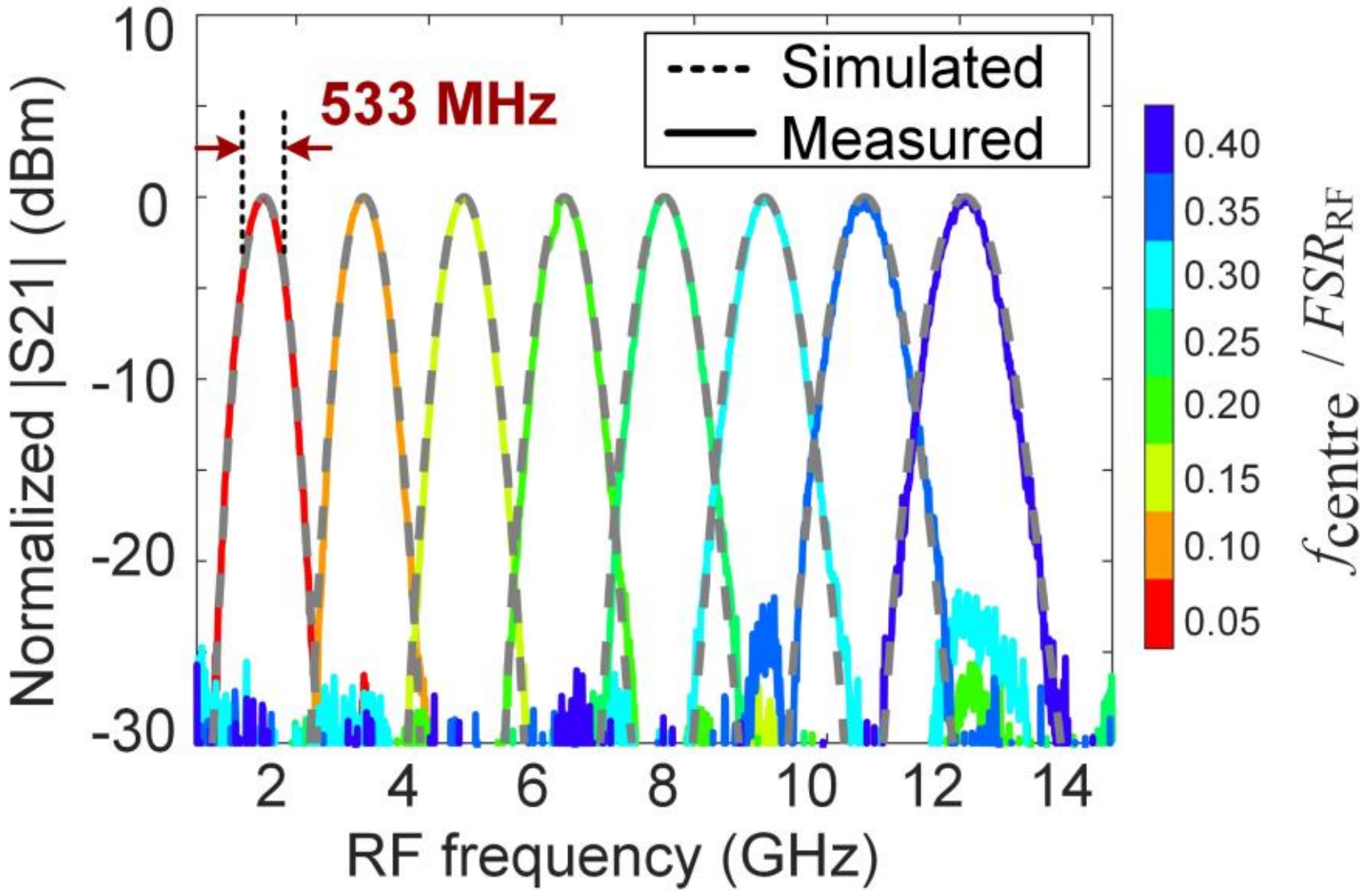

Fig. 9. Measured and simulated RF transmission spectra of the centre-frequency-tunable Gaussian apodized sinc filter

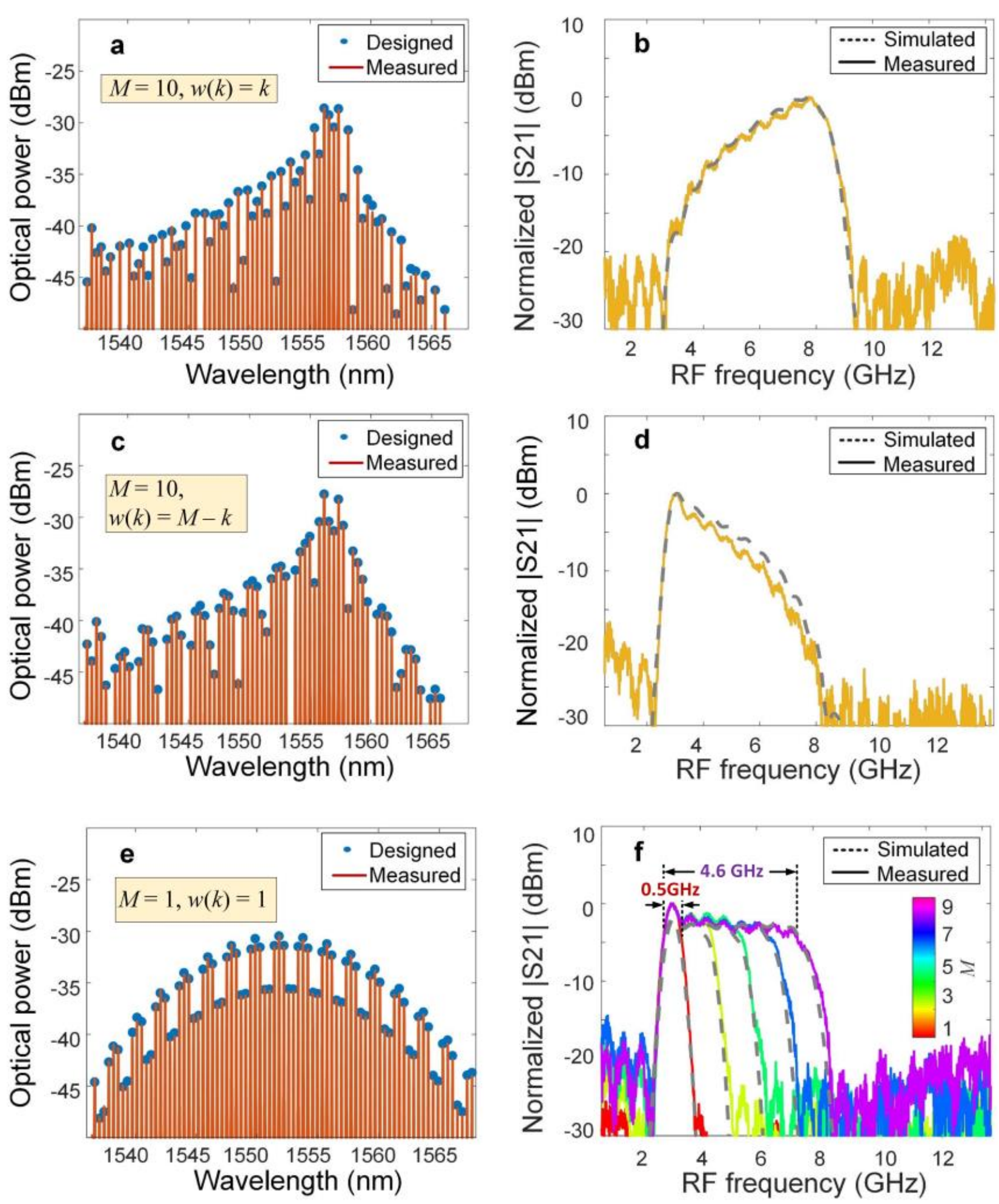

Fig. 10. Shaped micro-comb spectra (a, c, e) and RF transmission spectra (b, d, f) of the reconfigurable bandpass filter with (a, b) a positive transmission slope, (c, d) a negative transmission slope, and (e, f) varying bandwidth.

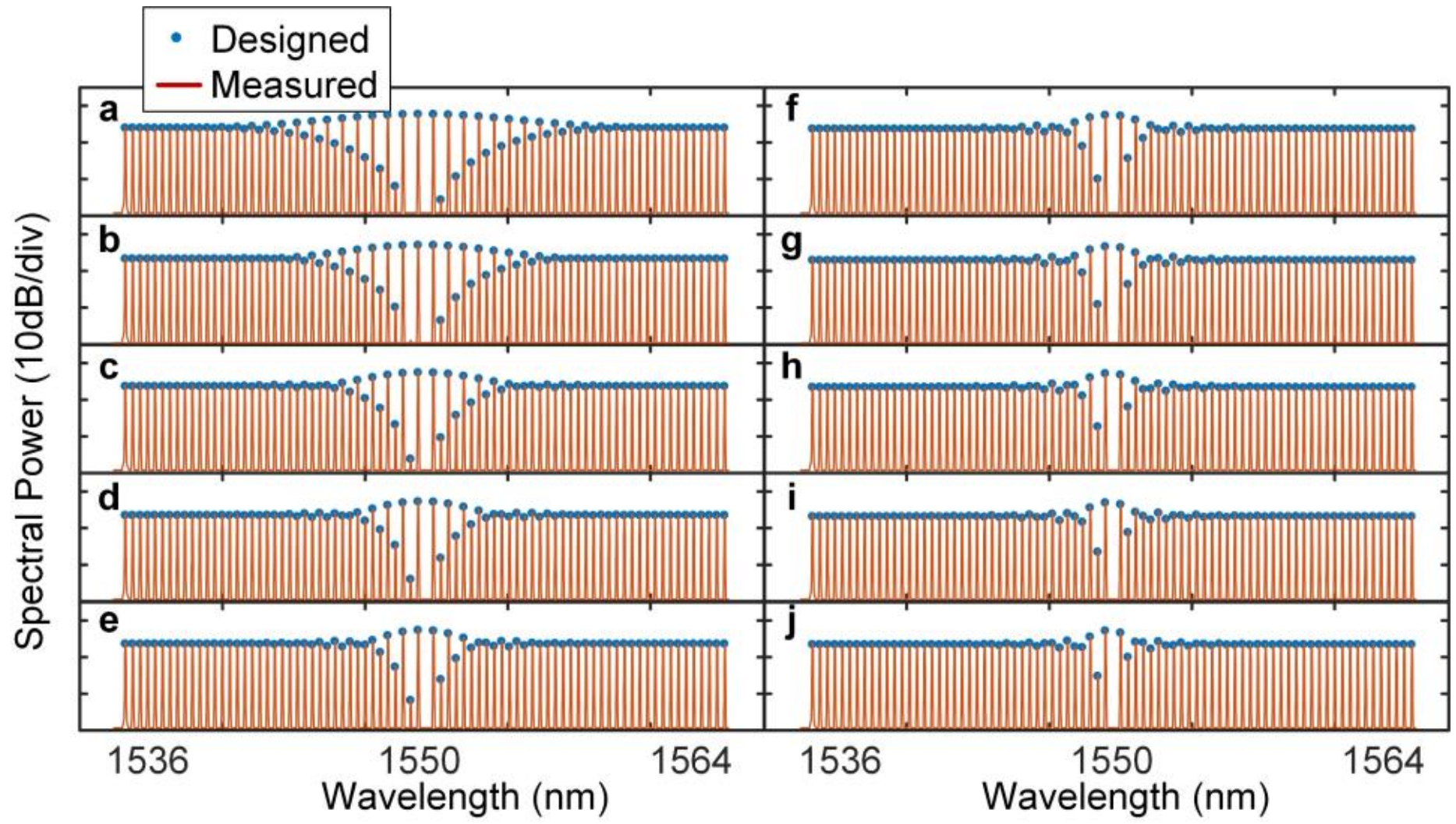

Fig.11. Optical spectra of the shaped micro-comb corresponding to the all-positive-tap bandwidth-tunable filter. (a-j) corresponds to a designed bandwidth from $0.025\times FSR_{\mathrm{RF}}$ to $0.25\times FSR_{\mathrm{RF}}$, with a step of $0.025\times FSR_{\mathrm{RF}}$.

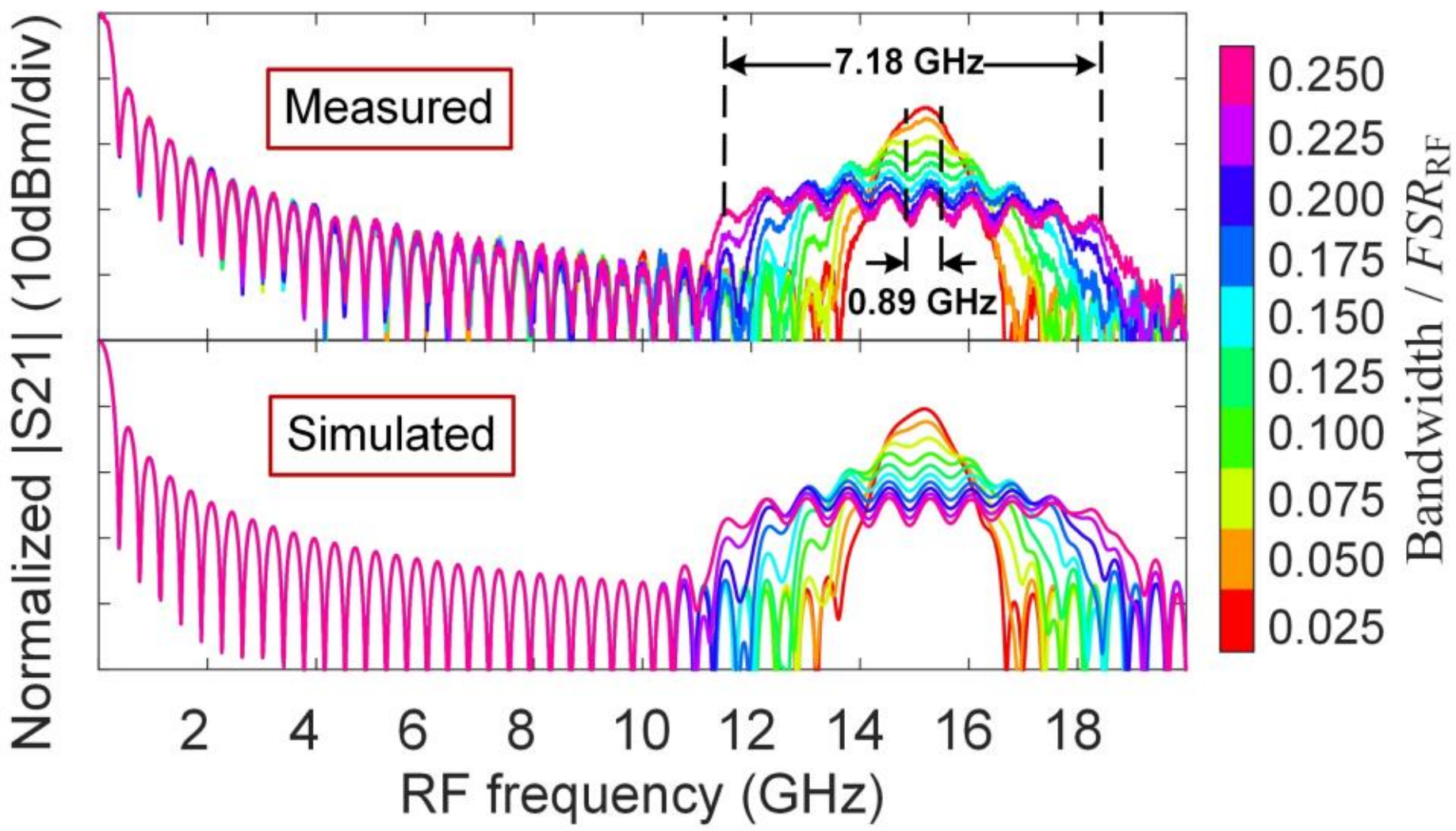

Fig. 12. Measured and simulated RF transmission spectra of the all-positive-tap bandwidth-tunable filter.